\documentclass[acmtog]{acmart} 
\acmSubmissionID{506}

\usepackage{booktabs}

\setcitestyle{nosort,square} 

\keywords{Geometric Deep Learning, Shape Synthesis}

\ccsdesc[500]{Computing methodologies~Neural networks}
\ccsdesc[500]{Computing methodologies~Shape analysis}

\usepackage{hyperref}
\usepackage{amsmath}
\usepackage{graphicx}
\usepackage{comment}
\usepackage{kbordermatrix}
\usepackage{multirow,bigdelim}

\usepackage{array}
\usepackage{wrapfig}

\usepackage{csvsimple}
\usepackage{adjustbox}
\usepackage[percent]{overpic}

\newcolumntype{C}[1]{>{\centering\let\newline\\\arraybackslash\hspace{0pt}}m{#1}}

\definecolor{olive}{rgb}{0.,0.55,0.27}

\newif\ifdraft
\drafttrue
\draftfalse

\newif\ifshowimages
\showimagestrue

\ifdraft
\newcommand{\dcc}[1]{{\color{red}[\textbf{DC:} #1]}}
\newcommand{\rhc}[1]{{\color{orange}[\textbf{RH:} #1]}}
\newcommand{\ahc}[1]{{\color{blue}[\textbf{AH:} #1]}}
\newcommand{\rgc}[1]{{\color{olive}[\textbf{RG:} #1]}}

\else
\newcommand{\dcc}[1]{}
\newcommand{\rhc}[1]{}
\newcommand{\rgc}[1]{}
\newcommand{\ahc}[1]{}

\showimagestrue
\fi

\ifshowimages
\newcommand{\imagepath}[1]{{#1}}
\else
\newcommand{\imagepath}[1]{{figures/99_dummy.png}}
\fi

\newcommand{\reals}{\mathbb{R}}

\newcommand{\rev}[1]{{\color{black}#1}}

\newcommand{\target}{target}
\newcommand{\source}{reference}
\acmJournal{TOG}
\acmYear{2020}\acmVolume{39}\acmNumber{4}\acmArticle{108}\acmMonth{7} \acmDOI{10.1145/3386569.3392471}
\begin{document}

\title{Deep Geometric Texture Synthesis}

\author{Amir Hertz}
\affiliation{\institution{Tel Aviv University}}
\authornote{Joint first authors.}

\author{Rana Hanocka}
\affiliation{\institution{Tel Aviv University}}
\authornotemark[1]

\author{Raja Giryes}
\affiliation{\institution{Tel Aviv University}}

\author{Daniel Cohen-Or}
\affiliation{\institution{Tel Aviv University}}

\begin{abstract}
Recently, deep generative adversarial networks for image generation have advanced rapidly; 
yet, only a small amount of research has focused on generative models for irregular structures, particularly meshes.
Nonetheless, mesh generation and synthesis remains a fundamental topic in computer graphics. 
In this work, we propose a novel framework for synthesizing geometric textures. It learns geometric texture statistics from local neighborhoods (\emph{i.e.,} local triangular patches) of a single reference 3D model. It learns deep features on the faces of the \rev{input} triangulation, which is used to subdivide and generate offsets across multiple scales, without parameterization of the reference or target mesh. Our network displaces mesh vertices in any direction 
(\emph{i.e.,} in the normal \textit{and} tangential direction), 
enabling synthesis of geometric textures, which cannot be expressed by a simple 2D displacement map. 
Learning and synthesizing on local geometric patches enables a genus-oblivious framework,
facilitating texture transfer between shapes of different genus.

\end{abstract}

\maketitle

\global\csname @topnum\endcsname 0
\global\csname @botnum\endcsname 0
\begin{figure}
    \includegraphics[trim={0cm 0cm 0cm 0cm},clip,width=\columnwidth]{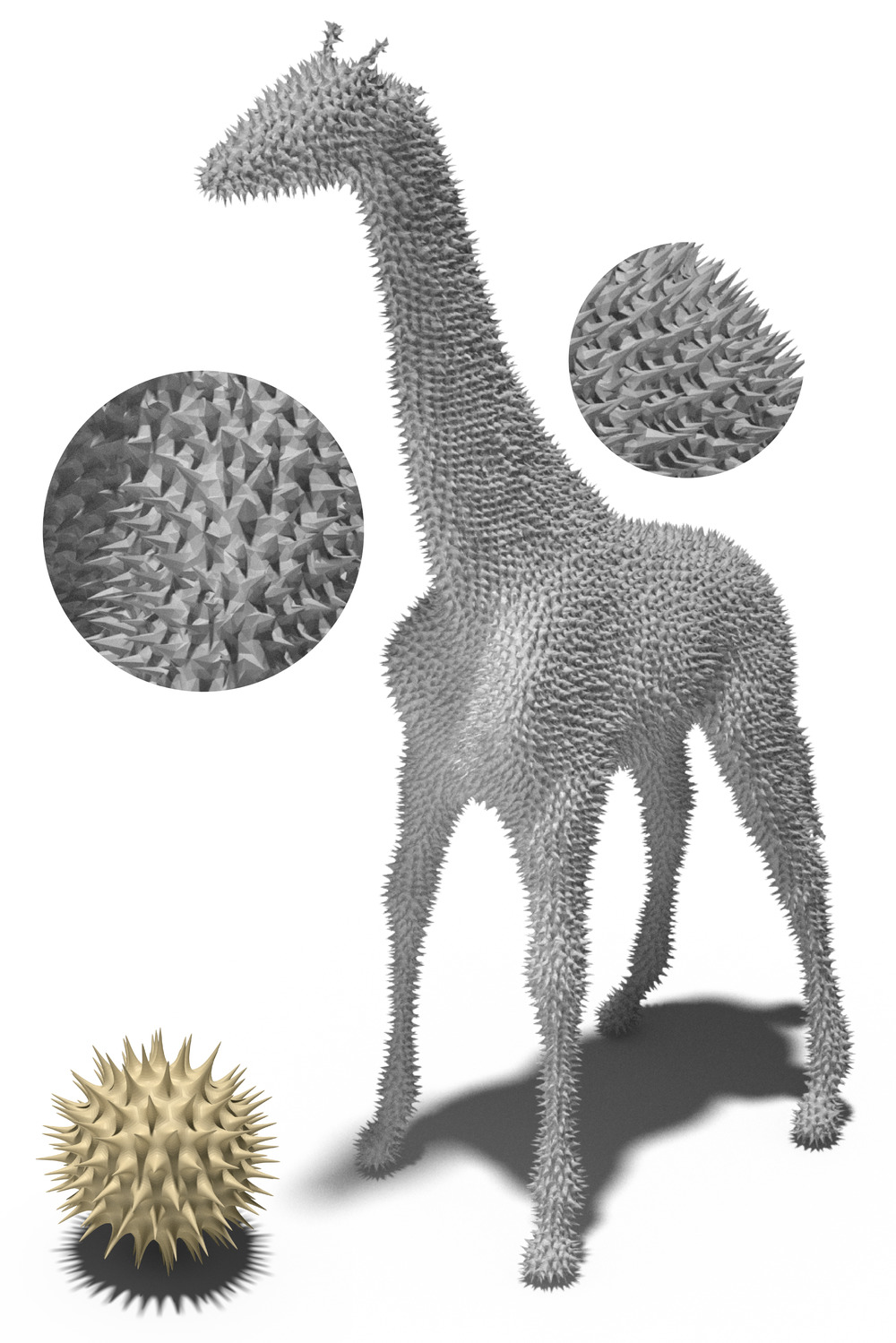}
    \caption{Learning local geometric textures from a reference mesh (gold) and transferring it to a target mesh (giraffe).}
    \label{fig:teaser}
    
\end{figure}

\section{Introduction}

In recent years, neural networks for geometry processing have emerged rapidly and changed the way we approach geometric problems. Yet, common 3D modeling representations are irregular and unordered, which challenges the straightforward adaptation from image-based techniques. Recent advances enable applying convolutional neural networks (CNNs) on irregular structures, like point clouds and meshes \cite{li2018pointcnn, Hanocka2019MeshCNN}. So far, these CNN-based methods have demonstrated promising success for discriminative tasks like classification and segmentation. On the other hand, only a small amount of research has focused on generative models for irregular structures, particularly meshes~\cite{gao2019sdm}.

In this work, we take a step forward in developing generative models for meshes. We present a deep neural network that learns the geometric texture of a single 3D \source{} mesh, and can transfer its texture to any arbitrary \target{} mesh. Our generative framework uses a CNN to learn to model the unknown distribution of geometric textures directly from an input triangular mesh.
Our network learns local neighborhoods (\emph{i.e.,} local triangular patches) from a \source{} model, which is used to subdivide and generate offsets over the \target{} mesh to match the local statistics of the \source{} model. For example, see Figure~\ref{fig:teaser}, where the geometric spikes of the \source{} 3D model are learned, and then synthesized on the \target{} surface of the giraffe.

\begin{figure*}[ht]
    \includegraphics[trim={0cm 0cm 0cm 0cm},clip,width=\textwidth]{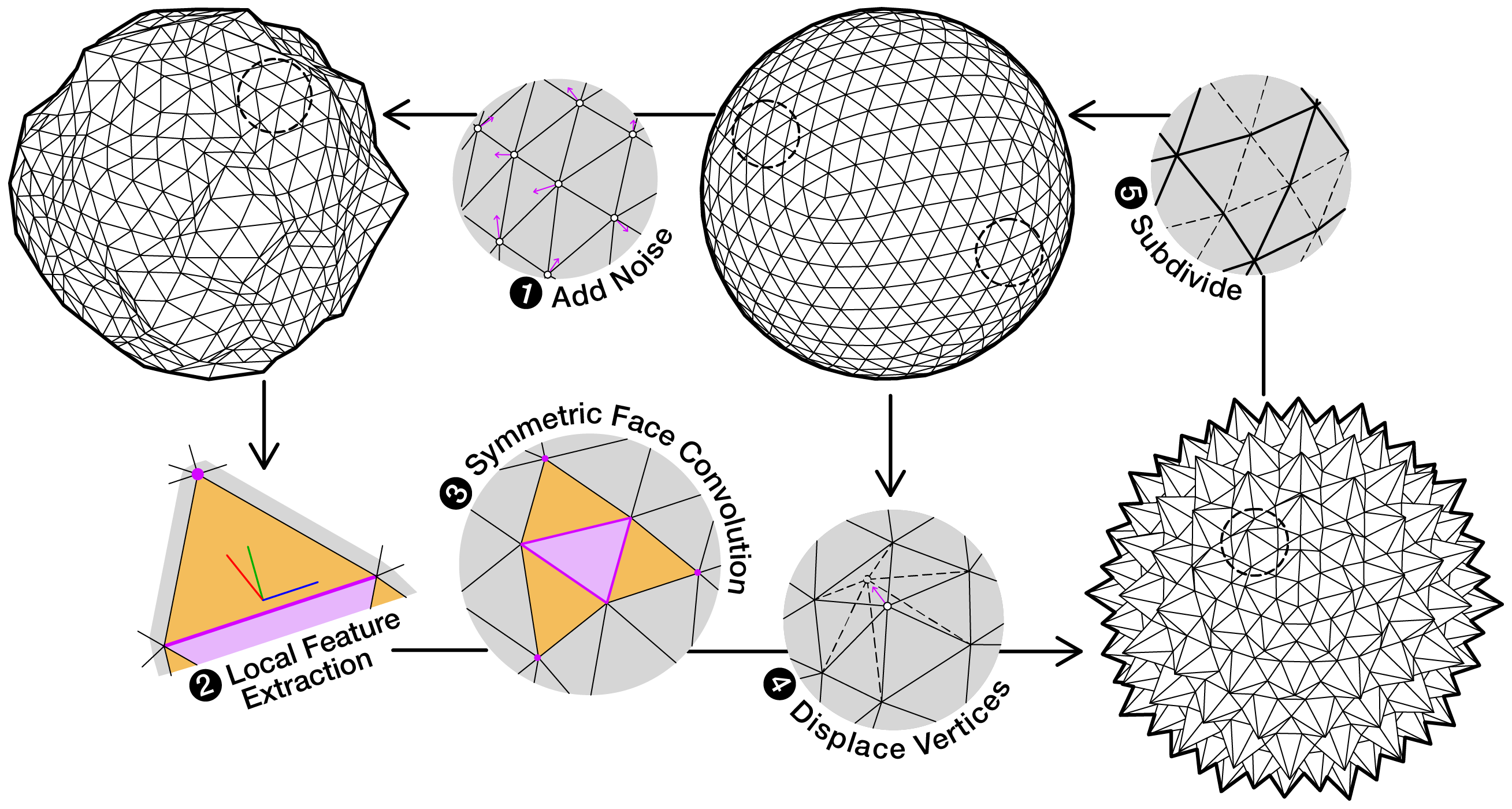}
    \caption{
    Method overview. Starting with the training input in the current scale in the hierarchy, we (1) add noise to the vertices and (2) extract local rotation and translation invariant input features per triangular face. We (3) learn face-based equivariant convolutional kernels which learn to generate differential displacements per vertex with respect to the input. Subdividing the generated mesh progresses to the next level in the hierarchy.}
    \label{fig:diagram}
\end{figure*}

In this work, we calculate deep features directly on the mesh triangles and exploit a unique property of triangular meshes. Every triangle in manifold triangular mesh is adjacent to exactly three faces (Figure~\ref{fig:diagram}), which defines a fixed-sized convolutional neighborhood, similar in spirit to MeshCNN \cite{Hanocka2019MeshCNN}. Our network generates mesh vertex displacements to synthesize local geometries, which are indistinguishable from the local statistics of the \source{} texture. To facilitate learning the statistics of geometric textures over multiple scales, we process the mesh using a hierarchy. We start with a low-resolution mesh (\emph{e.g.,} an icosahedron), and iteratively subdivide its faces and refine the geometry for each scale in the hierarchy.

Our method of transferring geometric texture from a \source{} model to a \target{} model has notable properties: (i) it requires no parameterization, of neither the \source{} nor \target{} surface; (ii) the \target{} surface can have an arbitrary genus, which is not necessarily compatible with the \source{} surface, and last but not least, (iii) it is generative: \source{} patches are not copied or mapped, instead, they are learned, and then probabilistically synthesized. Note that geometric textures can be rather complex, as shown in Figure~\ref{fig:tange}, which cannot simply be expressed by 2D displacement maps. Our network is given the freedom to displace mesh vertices in any direction, \textit{i.e.,} not only along the normal direction, but also tangentially.

We demonstrate results of transferring geometric textures from single meshes to a variety of \target{} meshes. We show that the \source{} mesh can have a different genus than the \target{} mesh. Moreover, we show that our generative probabilistic model synthesizes variations of the reference geometric texture based on different latent codes.

\begin{figure}[b]
     \includegraphics[trim={5cm 0cm 5cm 0cm},clip,width=\columnwidth]{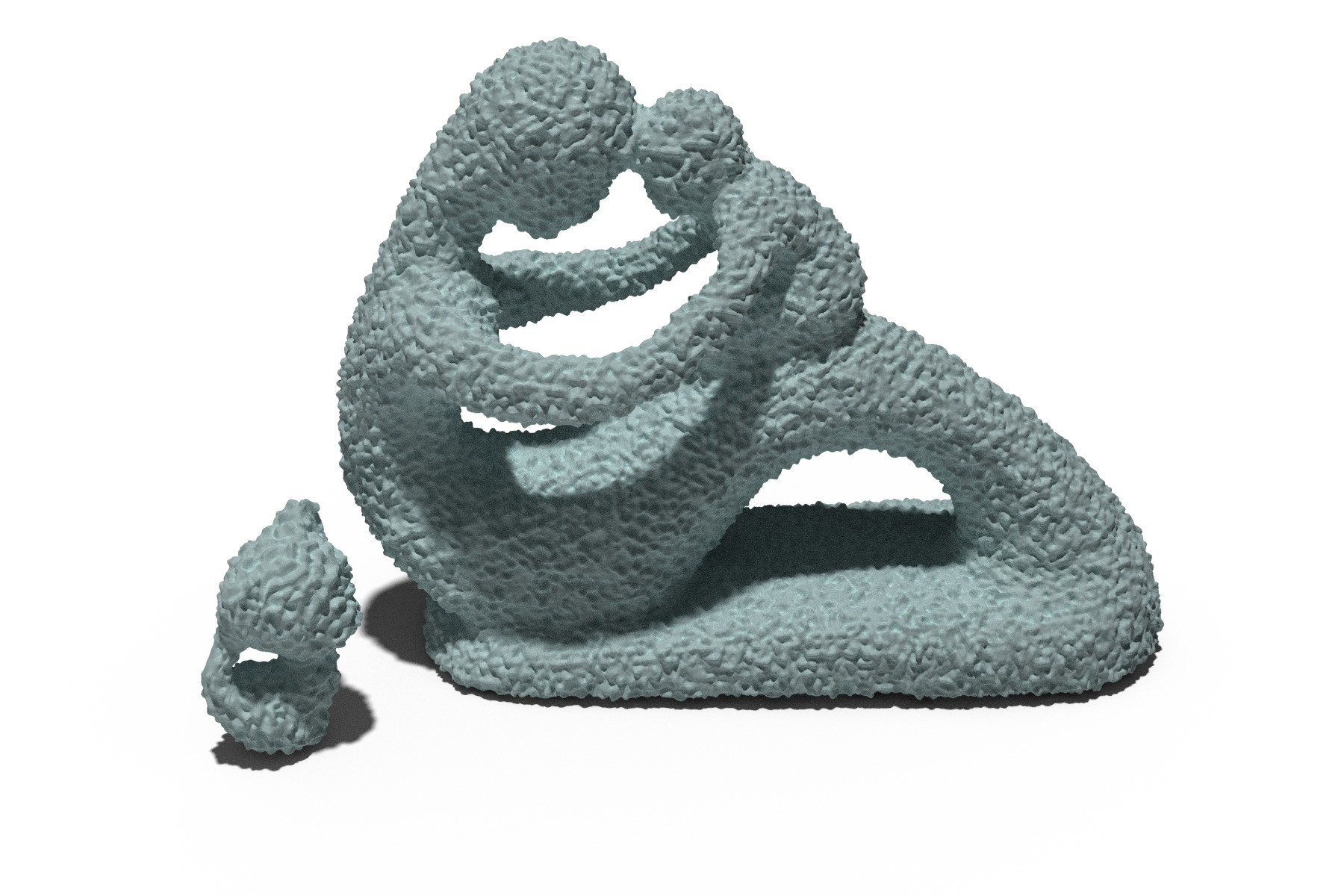}
    \caption{Our method is agnostic to the genus of both the reference and target meshes. Learning the geometric texture on a cat with a genus of one, and transferring it to the \textit{fertility} statue with a genus of four.}
    \label{fig:genus}
\end{figure}
\section{Related Work}

Generative models have garnered significant attention since the introduction of Generative Adversarial Network (GAN) \cite{goodfellow2014generative}. GANs are commonly trained on a large data set (typically images), attempting to generate novel samples that come from the distribution of the training data. Recently, some works presented GANs trained on a single image \cite{Zhou2018,shocher2018zero,DoubleDIP,shaham2019singan,sun2019test}. The basic idea is to learn the distribution of local patches from the patches of the reference image, and then apply the knowledge in various applications.
In the same spirit, in this work, we learn the distribution of local patches, but of 3D triangular meshes, which, unlike images, have an irregular structure. 

\textbf{Deep generative models in 3D.} In recent years, a large body of works have proposed generating or synthesizing 3D shapes using deep neural networks. 3D shapes are commonly represented by irregular structures, which challenge the use of traditional convolutional neural networks. Thus, early approaches proposed using a volumetric representation, which naturally extends 2D image CNN concepts to a 3D voxel grid~\cite{wu20153d, wu2016learning}. However, applying CNNs on 3D voxel-grids is highly inefficient, as it necessarily incurs huge  amounts of memory, particularly when a high resolution is required. 

On the other hand, a sparse and more direct portrayal of shapes uses the point cloud representation, which is simple and native to scanning devices. Achlioptas et al. ~\shortcite{achlioptas2018learning} pioneered the concept of deep generative models for point clouds, using the operators defined in pointnet~\cite{qi2017pointnet}, which uses $1 \times 1$ convolutions. Later, these ideas were extended to hierarchical structures and synthesis~\cite{qi2017pointnetpp, li2018so}. Recently, Yang et al.~\shortcite{yang2019pointflow} proposed a probabilistic framework for generating 3D point clouds. However, since the point cloud representation struggles to accurately portray fine grained details~\cite{hertz2020pointgmm}, it is not a common choice for representing shapes in 3D art or animation.

Alternatively, a 3D object surface and the fine grained details can be represented more accurately using some form of parameterization. Groueix et al. \shortcite{groueix2018atlasnet} propose representing surfaces as local charts, and use a deep network to learn a 2D parameterization for different tasks such as auto-encoders and surface reconstruction. The concept of using local charts for 3D shape generation has been further explored by \citet{ben2018multi}. \citet{kostrikov2018surface} propose generating surfaces through intrinsic networks. An alternative to an explicit surface is an implicit representation, for example using a signed-distance function and extracting the surface from its zero level-set. Recently, implicit fields have been proposed for 3D shape generation \cite{park2019deepsdf, chen2019learning, chen2019bspnet}.

The most common 3D representation in computer graphics is the polygonal mesh, a favorite of many due to the efficient, yet accurate portrayal of the underlying surface. Existing techniques for explicit mesh generation typically deform a template mesh, which preserves the genus and connectivity of the template.
Pixel2Mesh~\cite{wang2018pixel2mesh} is a network for generating genus-0 shapes by deforming a sphere from an input RGB image.
DIB-R \cite{chen2019learning3d} is a differentiable renderer which was demonstrated to reconstruct 3D objects from images. SDM-Net~\cite{gao2019sdm} is a VAE-based network for generating genus-0 mesh parts, yet, the collective sum of the parts can define non-genus zero shapes.

The most related work is MeshCNN~\cite{Hanocka2019MeshCNN}, a neural network with operators that delete and un-collapse edges from a mesh for discriminative tasks like segmentation. However, unlike \citet{Hanocka2019MeshCNN}, in this work, we propose a \textit{generative} network for synthesizing new mesh geometries. Since we learn from local geometric patches, our framework is oblivious to genus, and can transfer textures between arbitrary genus shapes (Figures \ref{fig:genus} and \ref{fig:noise}).

\textbf{Texture transfer on Meshes.} Texturing a target surface has been a fundamental problem in computer graphics. Basically, texture mapping requires parameterizing the target surface to define a low-distortion mapping between the source surface and target surface \cite{sorkine2002bounded, levy2002least, sheffer2007mesh}. In the most common setting, the source surface is a plane with a trivial parameterization. Naively, mapping a topological disc with boundary onto a manifold without boundaries, necessarily yields noticeable seams, where the boundaries are mapped and form discontinuities. Various works dealing with special textures with symmetries have developed continuous seamless mappings between closed surfaces (i.e., no boundaries) which have compatible genus \cite{aigerman2015seamless, aigerman2015orbifold, Knoppel:2015:SPS,campen2018seamless}.
\begin{figure*}[ht]
    \includegraphics[trim={0cm 0cm 0cm 0cm},clip,width=\textwidth]{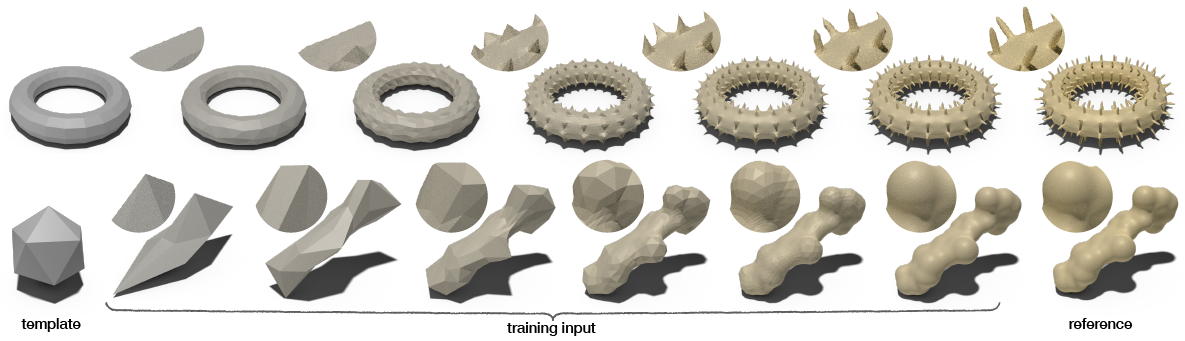}
    \caption{\rev{Multiscale training data generation. Given a reference mesh with geometric texture, we create a series of multi-resolution training data using an optimization strategy. Starting with a low-res template mesh we repeatedly subdivide and optimize the mesh geometry, to obtain a training input with increasing resolution.}}
    \label{fig:terminology}
\end{figure*}

Rather than mapping textures between surfaces, the textures can be synthesized over the target surface. Ying et al.~\shortcite{ying2001texture} and Turk et al.~\shortcite{turk2001texture} have presented texture synthesis techniques that synthesize textures from a 2D exemplar directly on the triangles of a target mesh. Their methods extend basic image space texture synthesis techniques by forming local parameterization over the mesh. Xu et al.~\shortcite{xu2009feature} present a more advanced method applied on meshes which is based on texture optimization \cite{wexler2004space, kwatra2005texture}. 

The above texture synthesis techniques are based on the premise that there is a simple local mapping between patches on the target and the source surfaces. Thus, they assume that the source surface is a flat image with a trivial parameterization~\cite{chen2012non}. The method we present does not map local patches, but learns the local geometries from the source mesh and synthesizes local geometries over the target mesh using a neural generative model. As noted earlier, local geometries are often too complex to be modeled by a simple $2D$ displacement maps. Recently, \citet{liu_2019} proposed an approach for cubic stylization, which can \emph{cubifiy} a 3D mesh directly, without any parameterization. Applying as-rigid-as-possible~\cite{sorkine2007rigid} reconstruction with an $\ell_1$ regularization on the normals leads to a cubic stylization that is detail-preserving.
\section{Deep Geometric Texture Synthesis}
\label{sec:method}
\begin{figure}
    \includegraphics[trim={0cm 0cm 0cm 0cm},clip,width=\columnwidth]{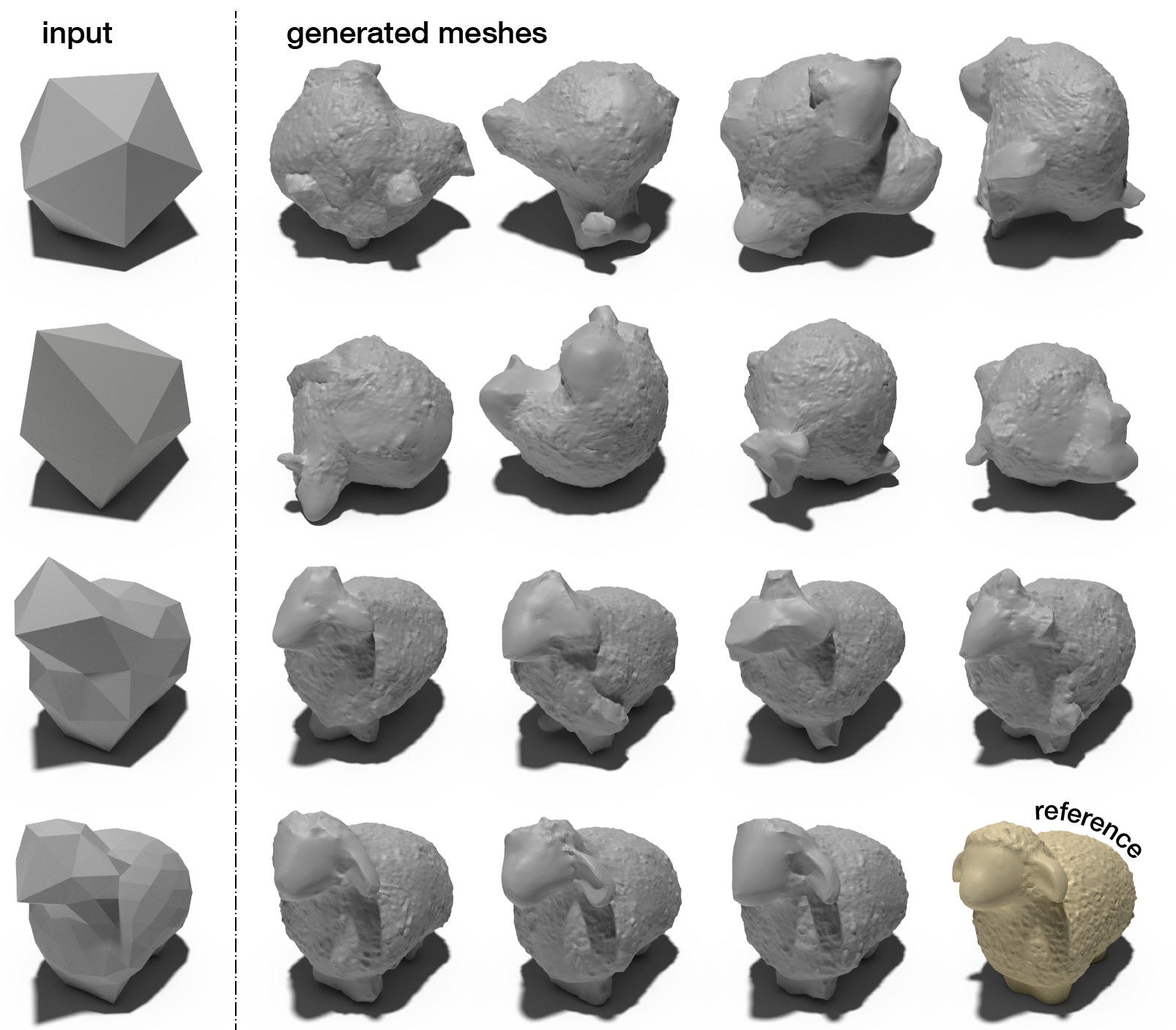}\\
    \includegraphics[trim={0cm 0cm 0cm 0cm},clip,width=\columnwidth]{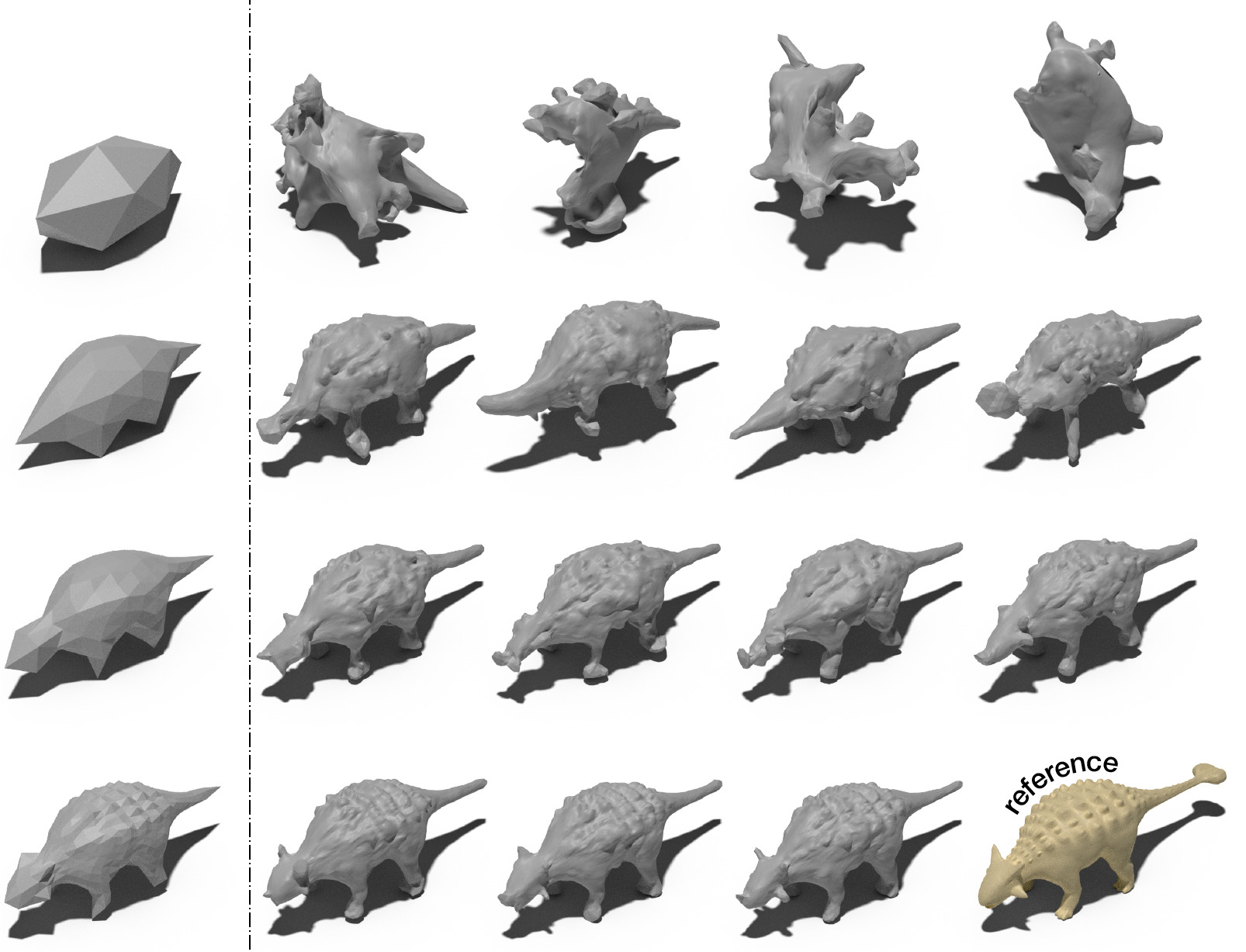}
    \caption{Unconditional mesh generation. Our method can unconditionally generate meshes (top rows), or conditionally generate meshes in different scale spaces. Higher levels in the scale space conditioned on a higher level input mesh results in a synthesis that maintains the global structure of the reference mesh.}
    \label{fig:generated}
\end{figure}
\begin{figure*}[ht]
    \includegraphics[trim={0cm 0cm 0cm 0cm},clip,width=\textwidth]{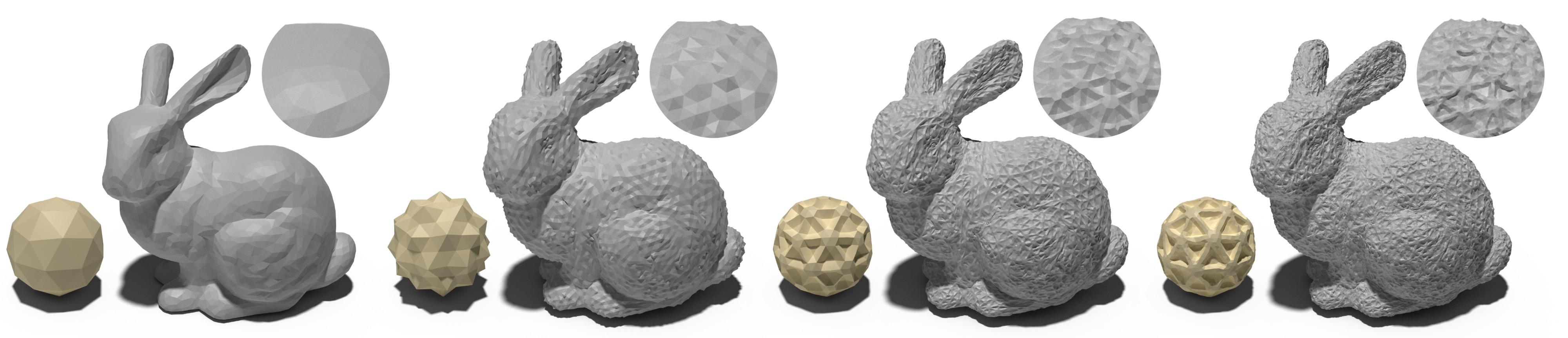}\\
     \includegraphics[trim={0cm 0cm 0cm 0cm},clip,width=\textwidth]{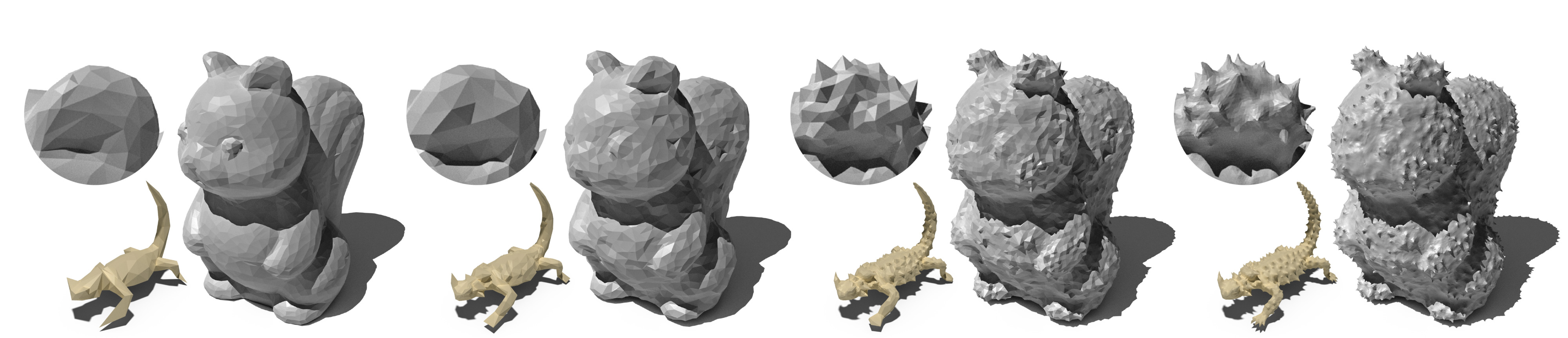}
    \caption{Hierarchical texture scale space. 
    A series of multi-scale generators are trained to synthesize geometric textures across multiple scales using the multi-scale training inputs (gold). During test time the geometric textures are synthesized on a novel target shape (gray). The scale space of the synthesized geometric texture is defined by the scale of the generators employed. The target shape is input to the first-level generator which synthesizes the first texture scale in the output (left). This output is passed to the second-level generator which synthesizes the next scale, and so on.}
    \label{fig:resolutions}
\end{figure*}
\begin{figure*}
    \includegraphics[trim={0cm 0cm 0cm 0cm},clip,width=\textwidth]{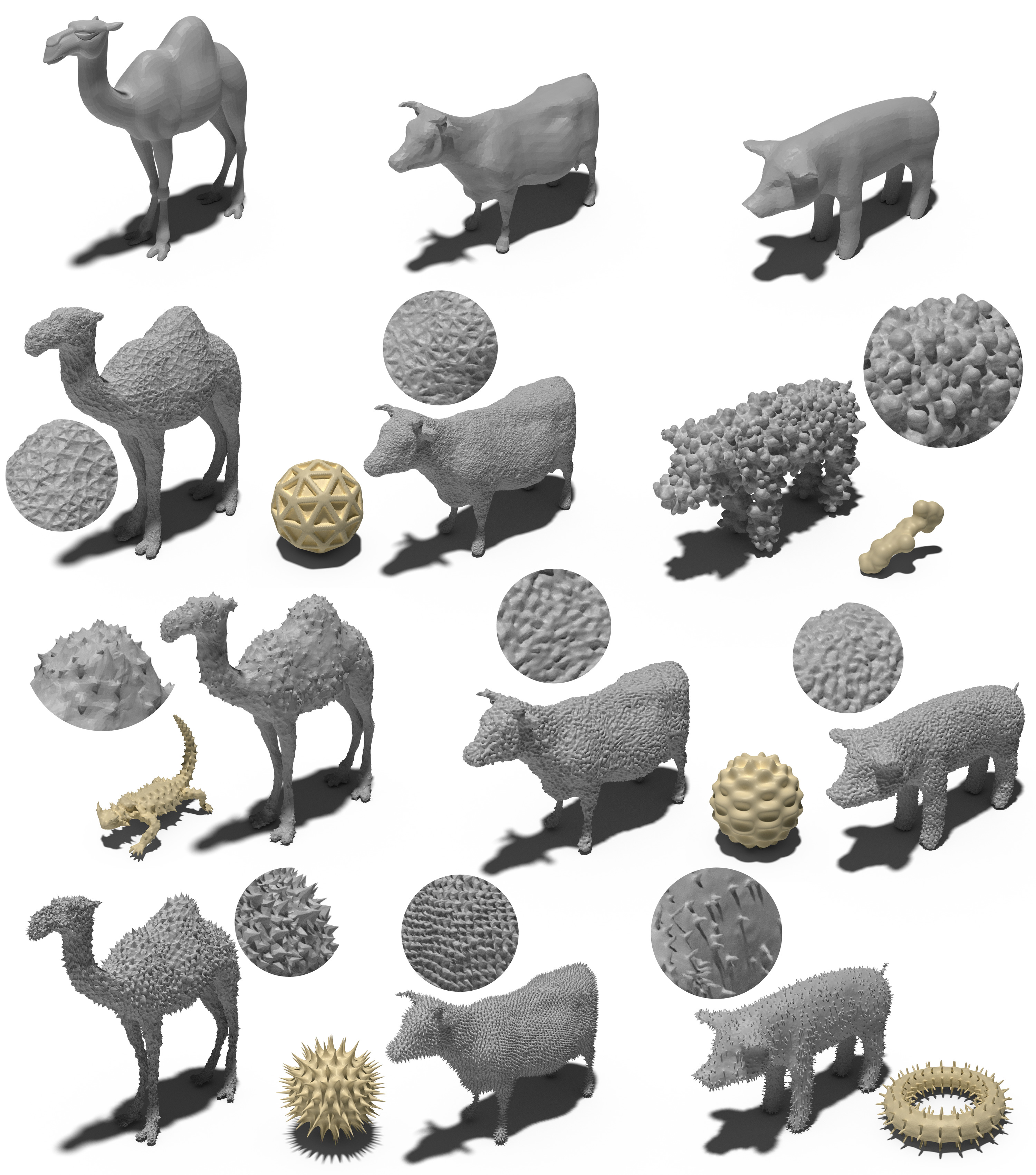}
    \caption{Geometric texture synthesis which are learned from a reference shape (gold) and transferred to different target shapes. Textures can be synthesized from natural shapes with geometric textures (\textit{e.g.,} the thorny lizard).}
    \label{fig:textures}
\end{figure*}

We present a framework for learning to synthesize the local geometric texture from a single mesh. 
We learn the statistics of the patches in a hierarchical, coarse-to-fine manner, where the input to each level is a subdivided version of the output coarser level. Figure \ref{fig:diagram} 
illustrates an overview of a single level in the hierarchy. We train a generative adversarial network (GAN) on the patches (\emph{i.e.,} local triangulations) of a single mesh, where the generator aims to synthesize local mesh patches that are indistinguishable from the reference patch.

Given a reference mesh with geometric textures, we create a series of meshes which depict the reference mesh across multiple resolutions. This multi-resolution series is used as input to train the hierarchical network. We obtain these \textit{multi-scale training inputs} via a preliminary optimization strategy. Starting with a low-resolution template mesh, the vertices are optimized such that its surface will match the reference mesh. This template is repeatedly subdivided and optimized to fit the reference, resulting in a multi-scale representation of the reference mesh (example in Figure~\ref{fig:terminology}). From this point forward, the reference mesh is \emph{discarded}, and these multi-scale training inputs are used to train the discriminator and generator.

Starting with the coarsest scale training mesh, we add Gaussian noise to its vertices, which are then used as input to the network. Then, we extract \textit{local} geometric features per triangular face, which are invariant to rigid transformations. The initial geometric features pass through a series of face convolutions to learn deep features. The output of the final convolutional layer is a displacement vector per triangular face, which describes a displacement for each of the three incident vertices. To generate the
final displacement vector per-vertex, we average the displacement vectors of all its incident faces. The displaced mesh is then refined by a subdivision and fed as input to the next level in the hierarchy.

The synthesized mesh in each scale is passed to the discriminator in the same scale, which learns to discriminate whether local patches (\emph{i.e.,} faces) are \textit{real} or \textit{fake}. The discriminator trains face-based convolutional kernels to abstract the input geometric features to salient deep features, which indicate whether the local mesh is synthesized or real. Note that the series of generators and discriminators have decreasing receptive fields that control the scale-space for synthesizing the geometric textures, in a similar hierarchical fashion as \citet{shaham2019singan} demonstrated on images.

After training is complete, we discard the discriminators, and use the series of multi-scale generators to displace vertices of \emph{any} novel target mesh. The scale space of the synthesized geometric texture is determined by the scale of the generators employed. See for example Figure~\ref{fig:resolutions}, where the training input (gold) is transferred to a target (gray) starting from the coarsest scale (left) to the finest scale (right).
Note that the target mesh may have a different triangulation and genus from the training input data.

In the process, we exploit a unique property of triangle meshes: the one-ring neighborhood of a mesh triangle has a fixed size of three triangles (step (3) in Figure~\ref{fig:diagram}). Similar to the edge-based convolution in MeshCNN~\cite{Hanocka2019MeshCNN}, we learn convolutional kernels which operate on the faces of each triangle, and the three neighboring faces of each triangle. We apply convolutions on the features of each face and the neighboring 1-ring faces. To be invariant to the initial ordering of the mesh, we apply symmetric functions to the features in the neighboring 1-ring, resulting in an equivariant triangular face convolution.

\subsection{Triangular Mesh Representation}
A triangular mesh is a special type of graph defined by a set of vertices and triangles: ($V$, $F$), where $V= \left \{ {v}_1, {v}_2 \cdots  {v}_n \right \}$ is the unordered set of vertex positions in $\mathbb{R}^3$. The mesh connectivity, or adjacency information, is designated by an unordered set of triangular faces $F=\left \{ {f}_1, {f}_2 \cdots  {f}_m \right \}$, each containing a triplet of vertices, which implicitly constructs the edges of the graph $E =\left \{ {e}_1, {e}_2 \cdots \right \}$ (pairs of vertices).

\textbf{Input Features.} At each resolution level, the input features to our network are defined locally on each face and describe the relations between each face and its three neighboring faces.
We define a local coordinate system for each edge in every face, where the origin is the edge midpoint. We use the face normal to define a consistent orientation for each local $x$, $y$, $z$ axis.
The local $z$-axis orientation is defined by the face normal, $x$-axis is the edge direction and $y$ is their cross product. Finally, we extract $4$ features for each edge: edge length and Cartesian coordinates of the opposite vertex to that edge, projected onto the local coordinate system (see step 2 of Figure~\ref{fig:diagram}). 

We denote the features of the three edges of the face $f$ by the matrix $S \in \reals^{3 \times 4}$,
where each row contains the features of a single edge.
These features are invariant to translations and rotations of the mesh. Moreover, these features contain enough information to reproduce the mesh in any global position and orientation from any face.

\subsection{Symmetric Face Convolution}
\label{sec:convolution}
In our network, we first perform a $1\times 1$ convolution on the input features to learn an order invariant \emph{face feature embedding}. Then, we perform a symmetric convolution that takes into account the three 1-ring neighbours of the face.

\textbf{Face Feature Embedding.} The geometric features per face serve as input features to our face-based convolutional neural network, which are subsequently abstracted to deep features. We denote the dimension of the feature vector in convolution layer $i$ by $d_i$. For the first convolutional layer of the network, we extract neural features $\hat{f} \in \reals^{d_1}$ for each face side via a linear layer $\hat{S} = g\left(S|W, b\right) = SW +b$, where $S \in \reals^{3X4}$ are the extracted features of that face, and $W \in R^{4 \times d_{1}}$ and $b \in R^{d_{1}}$ are learned weights. As we want to generate a face embedding that is invariant to the order of neighbouring faces, we apply a $max$ operation on the rows of $\hat{S}$. This leads to an initial face embedding $\hat{f}$ that is invariant to rigid transformations and mesh face order.

\textbf{Convolutions on Faces.} In the subsequent convolutions, the input face features are the deep embedding from the previous layer. Unlike the first block, in subsequent layers, the convolution operates on the face feature vector and the 3 neighboring face feature vectors.
Abusing notation, denote by $S \in \reals^{3 \times d_i}$ the matrix whose rows contains the intermediate face embedding of the neighbouring faces of a face $f$ and by $\hat{f} \in \reals^{d_i}$ its intermediate embedding. Then, we define the linear operation for the face by $g(S, \hat{f} | W_S, W_f, b) = SW_S + \hat{f}^{T}W_f +b$, where $W_S \in \reals^{d_i \times d_{i+1}}$ , $W_f \in \reals^{d_i \times d_{i+1}}, b \in{\reals^{d_{i+1}}}$ are the learned weights. To ensure the convolution is invariant to the face ordering, we take a $Max$ operation across the rows of $g(S, \hat{f} | W_S, W_f, b)$.

\textbf{Vertex displacement.} In this work, the face-based convolutions are used to build both the discriminator and generator networks. The discriminator uses the deep feature embedding to distinguish between \textit{real} and \textit{fake} faces, while the generator outputs $3D$ displacement vectors which modify the input mesh geometry.
The generator outputs a single displacement vector per face, which is used to displace its three vertices symmetrically. 

Each face predicts a displacement vector that is shared across all three vertices in that face, which is then projected onto the local coordinate axis of each edge, respectively. Since each vertex is shared by several faces, it receives multiple displacements. We average all of them to calculate its final displacement. Note that while each face predicts a symmetric displacement to each vertex, the vertices can be moved in all directions since they receive displacements from all the incident faces.

\subsection{Subdivision and Multi-Scale Meshes}
\label{sec:subdivision}
Realizing our goal to learn to synthesize geometric textures from a reference mesh, requires defining a method for \emph{upsampling}, or subdividing the input mesh to achieve a hierarchical scale space. After defining a subdivision operator, it will be used to iteratively increase the resolution of some input mesh, such that with each subdivision, additional details from the reference mesh are added to the input mesh (see golden mesh in Figure~\ref{fig:resolutions}).

\textbf{Uniform Subdivision.}
In images upsampling is trivial, since downsampling and upsampling results in the same connectivity, \emph{i.e.,} the local scale space of the image grid is preserved. 
However, for the irregular mesh structure, we must define an operator which can upsample both the training and inference meshes in a similar manner. For example, it is not sufficient to simply collapse edges in some pre-defined order, and then restore the collapses, since there would be no way of transferring this operation to a novel mesh. To this end, we propose using a uniform subdivision operator, which will have the same behavior on any given connectivity. 

Uniform subdivision divides each face into four faces, by placing a triangle inside each face (see step 5 of Figure~\ref{fig:diagram}). A vertex is placed in the midpoint of every edge in the triangle, which increases the mesh resolution by four. This operation is \textit{fixed}, meaning that given a specific connectivity (regardless of vertex placement), the uniform subdivision will always generate the same mesh. This property is desirable for transferring to novel \target{} meshes which have a different connectivity than the training mesh.

\textbf{Multiscale Input Shapes.} Given a \source{} mesh with geometric texture, we employ a pre-processing phase to prepare a series of multi-scale input shapes which we will use for training. The user defines a low resolution template which is iteratively subdivided and deformed to match the \source{} mesh. The template was chosen to be either: an icosahedron, torus or coarse mesh (simplified version of the reference). Note that for a given \source{} mesh, the exact tessellation will not be recovered using uniform subdivision from some template. For this reason, we \emph{remesh} the reference shape prior to training, and only use the multi-scale inputs during training (\emph{i.e.,} discard the reference).

The proposed re-meshing procedure will generate a series of multi-scale training inputs. We create increasing resolutions (or scales) of the reference mesh via an optimization procedure. Starting with a template mesh, we iteratively subdivide and minimize the distance to the reference mesh. As the number of mesh elements increase, the optimization will obtain a better fit to the reference mesh.

We solve this optimization problem through back-propagation, where the minimizer is the vertex locations of the training meshes. 
The optimization objective is measured by a bi-directional Chamfer distance between uniformly sampled points on both the reference and the optimized mesh. This distance is the Euclidean distance between each point on the training mesh and its closest point in the reference (and vice-versa), in addition to a negative cosine similarity between the normals on the meshes at those points.

We add two regularization terms to this optimization process to obtain a locally uniform triangulation and a smooth shape. The first (uniform) term encourages the minimization of the variance of the edge lengths and the second (smoothness) term reduces the distance between each vertex $v_i$ on the mesh to the average coordinate of its one-ring: $$\left|v - \dfrac{1}{d_{i}}\sum\limits_{j:\left(i, j\right) \in E} v_{j}\right|,$$ where $d_i$ is the degree of the vertex $v$.

\begin{figure}
    \includegraphics[trim={0cm 0cm 0cm 0cm},clip,width=\columnwidth]{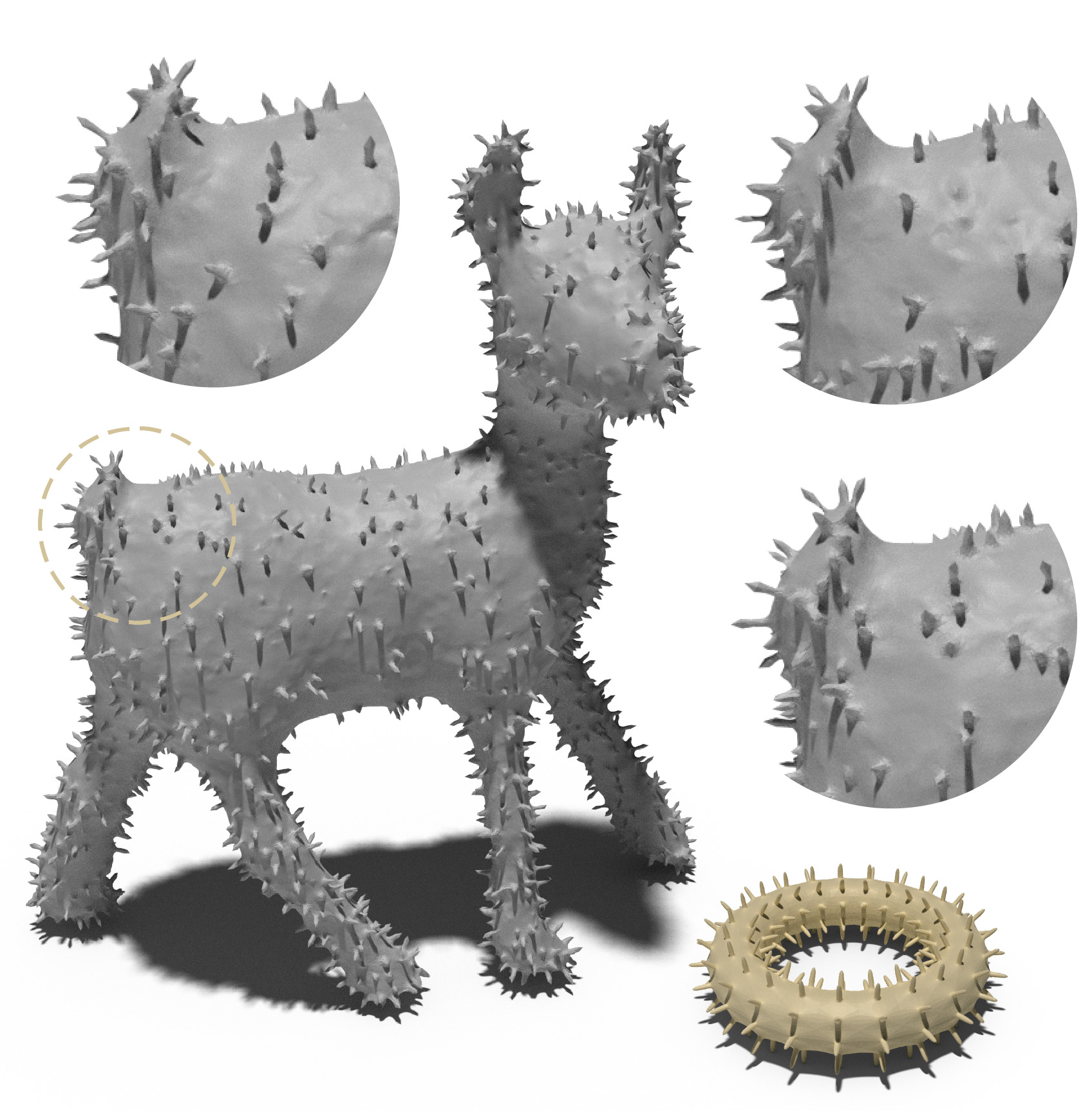}\\
    \caption{Latent synthesized textures. By sampling different noise vectors, we can synthesize \rev{variations of the geometric texture.}}
    \label{fig:noise}
\end{figure}

\subsection{Learning from a Single Mesh}
\label{sec:learning}
We describe how we use our face-based convolutional layers to design a GAN model (generator and discriminator) that learns the local geometric statistics from a single mesh using a hierarchy of generators. 
The generator network learns to predict vertex displacements to generate local geometries which are indistinguishable from the local statistics of the reference texture.

\textbf{Hierarchical GAN training.} We synthesize geometric textures via a series of generators which create local geometries incrementally. The output of a generator at a given level is local refinements to the input mesh, which is subdivided and used as input to the generator in the next level. In this manner, displacements in the coarse generator correspond to large refinements on the final mesh, and as the mesh progresses through the hierarchy, the generator displacements become fine-grained. This eases the training since each generator level only needs to capture the local refinements of each scale.

The generator in our model receives an input mesh and a tensor of noise $z$ that is added to the input mesh vertices. Then, the generator outputs a displacement vector per-face, which is applied on the input mesh shape (without noise). The discriminator receives both the modified input mesh (\textit{fake}) and the corresponding \textit{real} mesh (\emph{i.e.,} training shape with the same resolution) as input. An illustration of the \textit{real} meshes for each level is shown in Figure~\ref{fig:terminology}.
The discriminator is patch-based, so it learns to classify whether faces are \textit{real} or \textit{fake}. In other words, given an input mesh, the discriminator estimates a probability per face of being \textit{real}.

As with standard GAN training, the goal of the generator is to fool the discriminator by generating shapes that are as similar to the \textit{real} mesh, and the goal of the discriminator is to be able to distinguish between the generator output and the true mesh. We use the WGAN-GP \cite{gulrajani2017improved} framework to train both. 

In addition to the adversarial loss, we add also a reconstruction loss as suggested in~\cite{shaham2019singan}.
We require that for a given fixed noise vector $z = c$, the generator will be able to reconstruct the \textit{real} mesh. We use the $MSE$ distance between the vertices of the generated and real meshes. To combine the two loss functions, we use a parameter $\gamma$ to weight the reconstruction loss. 

Training starts with the generator and discriminator in the coarsest level. The input to the coarsest generator is the template (+ noise $z=c$), while the \textit{desired output} (\emph{i.e.,} using reconstruction and adversarial loss) is the \textit{training input} in the coarsest level. Both the generator and the discriminator are trained until convergence. When progressing to train the next level, the generator from the previous level is kept fixed. The output from the previous level is subdivided and scaled and then input to the next level. After subdivision, we uniformly scale the mesh such that the mean-edge length is preserved. We set $c$ to be a fixed random noise vector in the coarsest resolution, and a vector of zeros in the higher resolutions.

\textbf{Inference.} Our generator network is fully convolutional, and therefore it can be applied to any mesh with any connectivity and resolution. Given a new shape (\emph{i.e.,} \textit{target} mesh), we use the generator to synthesize the learned local structures of the reference mesh. This is achieved by scaling the target shape to have an average edge length of one (\textit{i.e.,} input feature normalization). We use the target shape plus random noise, as an input to the generator in one of the lower resolutions in the hierarchy. This leads to transferring the (local) structure of the reference mesh to the target mesh. Note, unlike the reference mesh, the given connectivity of the target mesh does not need to be re-meshed.
\section{Experiments} 
\label{sec:experiments}
\begin{figure}
    \includegraphics[width=\columnwidth]{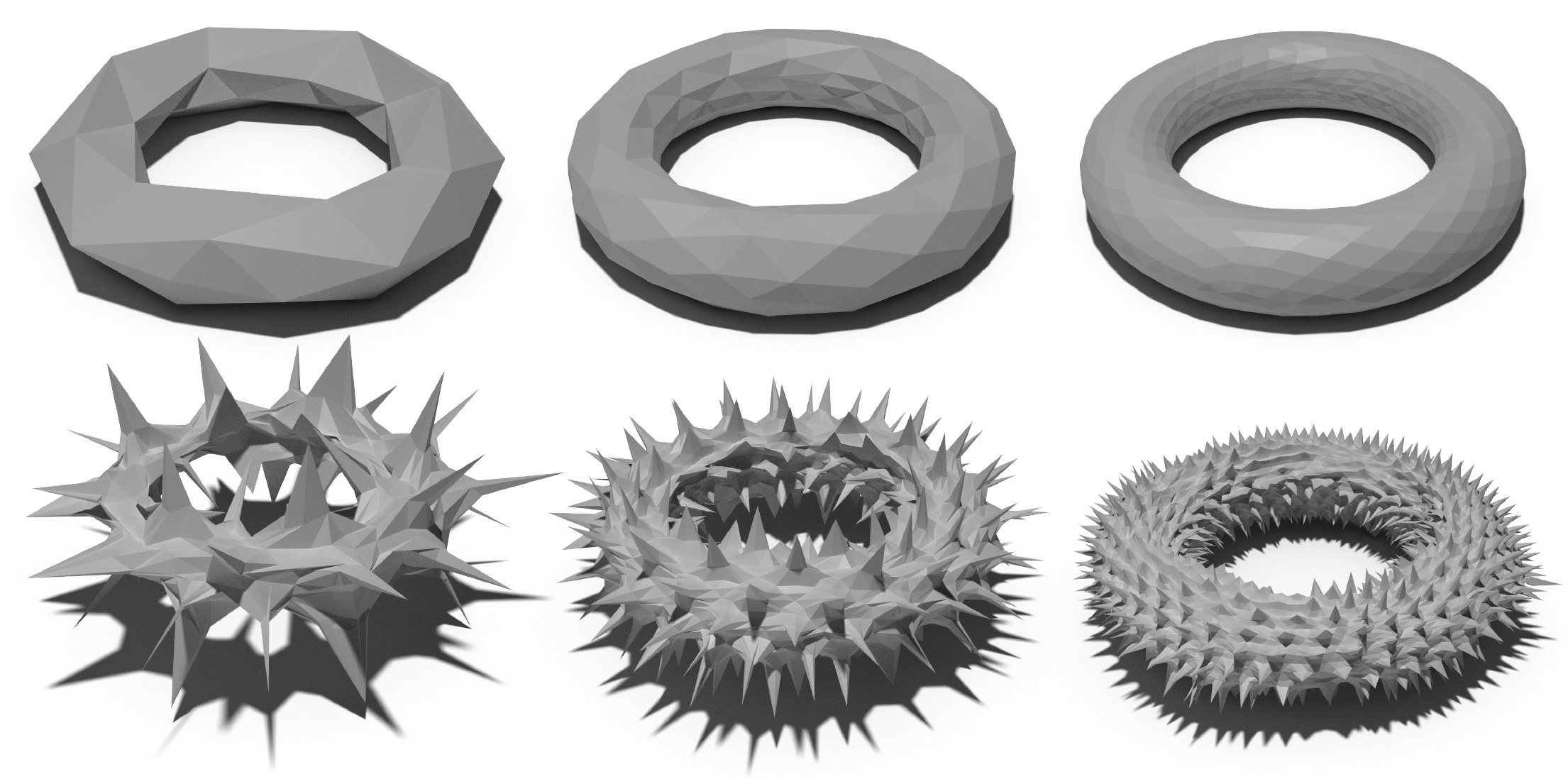}
    \caption{Transferring geometric texture from the spikey ball (shown in Figure~\ref{fig:teaser}) to a torus with different resolutions. Transferring the spikes to a \textit{low} resolution torus results in coarse texture scale space. Increasing the resolution of the torus increases the transferred texture scale space.}
    \label{fig:torus}
\end{figure}
\begin{figure*}
    \includegraphics[trim={0cm 0cm 0cm 0cm},clip,width=\textwidth]{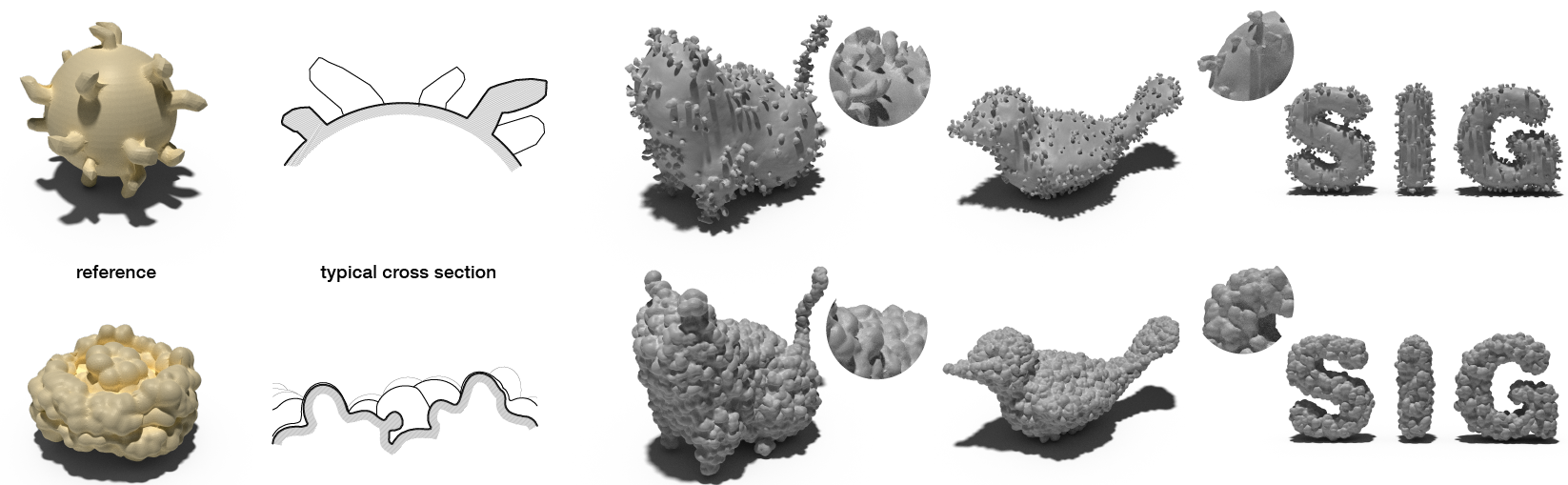}\\
    \caption{Geometric textures with complex 3D displacements. The network learns to synthesize geometric textures from the reference geometry, whose tangential movement is highlighted in a cross section illustration, respectively. Synthesizing the geometric textures on different \target{} shapes (right - gray).}
    \label{fig:tange}
\end{figure*}

The reference meshes using for training our models are provided by Thingi10K dataset \cite{Thingi10K} or built by hand. Our PyTorch \cite{paszke2017automatic} implementation as well as pre-trained models and multiscale training meshes will be made publicly available upon publication.

Each model contains $5-7$ levels (scales), depending on the complexity of the geometric texture in the reference mesh.
Across all levels, the generator and discriminator have $7$ layers of face convolutions with instance normalization \cite{ulyanov2016instance} and leaky ReLU.
The face feature embedding dimension ($d_1$ in Section~\ref{sec:convolution}), or the output of the first convolution layer increases as we move up the hierarchy. Starting with 32 in the first level, and reaching 128 at the third level.
From the fourth level and onward, we initialize both the generator and the discriminator models, with the weights of the previous level.
Each level was trained for $2000$ iterations using the Adam optimizer \cite{kingma2014adam} with a learning rate of $5e^{-4}$ and learning rate decay of $0.5$ applied in intervals of $500$ iterations. In all experiments, the reconstruction weight $\gamma$ was set to 5.

\textbf{Hierarchical Generation.} Our hierarchical training allows synthesizing meshes starting from different levels of the generator.
When starting from the lowest level with the template, the generator outputs different meshes with different global structures (see top rows of Figure \ref{fig:generated}).
When synthesizing from higher levels by using higher resolution inputs, the generator preserves the global structure of the input and only deforms local regions on the mesh (see bottom rows in Figure \ref{fig:generated}).

This hierarchical characteristic enables applying our model on a variety of meshes with different resolutions during inference from any given level. In general, we usually start from level 2-3 of the source shape when performing the texture synthesis. However, this is dependent on the training meshes, and the scale spaces that are defined within each level. Some geometric textures require more levels, while others can be compactly defined in a few levels.

We evaluate the pretrained models by applying them on unseen target meshes. Figure~\ref{fig:resolutions} shows the hierarchical  generation of textures. Notice how the geometric texture is transferred gradually from the source shape (in gold) to the target shape, where the process starts using the source shape in a lower resolution and progresses while increasing it.

\textbf{Texture Synthesis.} Figure \ref{fig:textures} presents additional texture synthesis for various unseen target meshes from different reference shapes. Observe how our method is able to synthesize different geometrical structures on directly on the target shape, without the use of any parametrization. 

A remarkable property of our approach is that it can process pairs with a different genus. See for example the torus and the pig in Figure~\ref{fig:textures}. In Figure~\ref{fig:genus} the generator was trained on the cat model (genus of one), and was transferred to a target fertility shape with a genus of four. 

Notice that the resolution (number of mesh faces) used in the target shape determines the scale of the texture synthesized on them. Figure~\ref{fig:torus} demonstrates this effect by synthesizing spikes on a torus mesh, where we start from different mesh resolutions and transfer the texture from levels $2-4$ (top row). In all three cases, we synthesized the textures using the same generator scales trained on the spikey ball. Naturally, the resolution of the target affects the size of the synthesized texture.

\textbf{Latent Space Interpolation.} Since our framework is generative, it enables synthesizing different textures from the same reference shape. This can be done by 
sampling different noise vectors, resulting in different synthesized textures on the target mesh. We show examples on two different shapes and textures in Figure \ref{fig:noise}. 
Note that since the generator was trained on a single reference mesh, the differences in the synthesized texture on the same target shape are solely due the noise vector. This enables smoothly interpolating between shapes by simply interpolating over the latent variable that was used for generation. Performing smooth interpolations between shapes enables animation of the textures. We provide several such examples in the supplementary material.

\begin{figure}
    \includegraphics[trim={0cm 0cm 0cm 0cm},clip,width=\columnwidth]{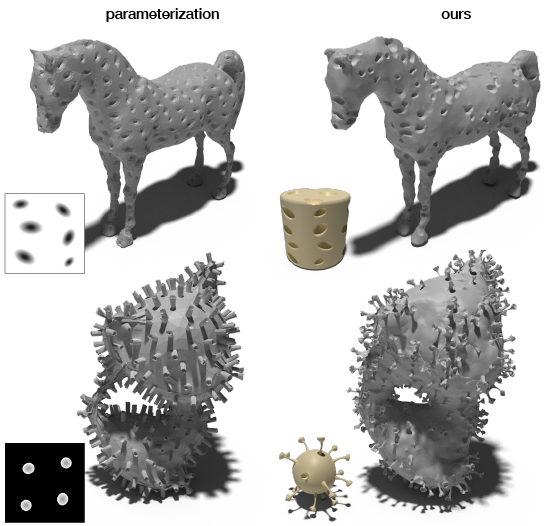}\\
    \caption{Comparison to OptCuts ~\cite{Li:2018:OptCuts} (left), which projects a 2D displacement map to the target surface. This UV mapping is used to displace vertices in the normal direction. 
    The 2D displacement was estimated manually from the 3D reference shapes in gold: cylinder and coronavirus. Note the tangential displacements of the coronavirus are not captured by a 2D displacement map.
    Our technique (right) learns to synthesize 3D geometric textures directly from the reference mesh (gold).}
    \label{fig:comparison}
\end{figure}

\textbf{Comparison.} We compare against OptCuts~\cite{Li:2018:OptCuts}, a state-of-the-art parameterization technique, in Figure~\ref{fig:comparison}. We manually create a 2D displacement map which corresponds to the 3D reference shape (gold), and use OptCuts to automatically calculate the parameterization and cutting of the 3D mesh, resulting in a mapping of the displacement map to the target mesh. We use the UV mapping to displace vertices in the normal direction on the target mesh. 
On the other hand, our technique learns to synthesize geometric textures directly from a 3D reference mesh. Finally, it can be seen that the edges of OptCuts textures are sharp. However, automatically creating 2D displacement maps from 3D geometries is non-trivial. Moreover, a 2D displacement map that moves geometry in the normal direction is rather limited, since it does not encode tangential movement (\textit{e.g.,} the coronavirus). Lastly, OptCuts took 10 minutes to compute a parameterization, while our technique only requires a few seconds.

\begin{figure}[b]
     \includegraphics[trim={0cm 0cm 0cm 0cm},clip,width=\columnwidth]{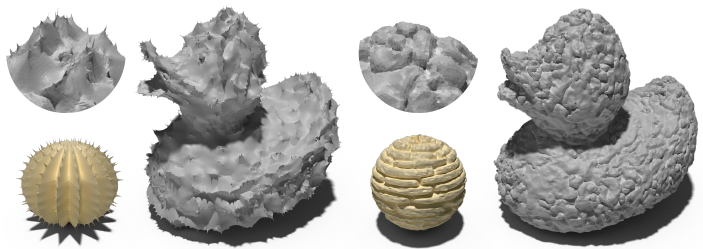}
    \caption{Our method is limited to isometric textures. The vertical (cactus) and horizontal (brick) texture direction is not transferred to to the duck.}
    \label{fig:limitation}
\end{figure}

\section{Discussion and Future Work}

We have presented a novel concept for geometric texture synthesis, which uses a generative framework to learn the local structures from a given triangular mesh and then synthesizes it on different target models. Our technique learns to match the local statistics of a specified mesh model and transfers it to a target one. To the best of our knowledge, this is the first generative model that learns from a single mesh. 

A prominent advantage of our scheme is that it does not require any parameterization of the reference or target shape. Given a model with natural organic geometric texture (i.e., lizard Figure~\ref{fig:resolutions}), which is not given as a displacement map, it is not immediately obvious how to employ a classic parameterization technique to transfer the geometric texture to another (target) shape (\textit{i.e.,} squirrel Figure~\ref{fig:resolutions}). There is no generic method for decomposing an arbitrary surface with geometric textures into a \textit{base} and displacements. Furthermore, not every geometric texture is simple enough to be represented as displacements along the surface normal (\emph{e.g.,} reference shapes with tangential movement in Figure~\ref{fig:tange}). By contrast, our approach receives a reference 3D model which contains geometric texture (\textit{i.e.,} not a displacement map), and learns to synthesize geometric structures by displacing vertices in all directions (\emph{e.g.,} not only along the normal direction but also tangentially).

However, the presented method has its limitations. First, it learns to synthesize local textures, and cannot capture large structures. Moreover, it currently assumes that the geometric textures are stationary and isometric (\emph{e.g.,} Figure~\ref{fig:limitation}). 
Handling anistropic textures would entail learning a directional field which can be transferred from the reference to the target mesh, a difficult task in and of itself. Moreover, even after the directional field is estimated, synthesizing the final geometric texture from the directional field is not a trivial task.
Another limitation is that the hierarchical learning requires the mesh to have a locally-uniform triangulation and well-behaved subdivision structure. Currently, we achieve this via a preliminary remeshing process. This remeshing procedure may fail on complex shapes, \emph{e.g.,} thin and intertwined structures, and in general, where the euclidean and geodesic distances differ significantly. In the future we would like to relax this requirement, and build the hierarchy by learning vertex splits.

While the focus of this work is geometric texture synthesis from a single mesh, our approach opens the door for a variety of follow up works. For example, it is possible to use the machinery developed in this work for transferring color texture or other attributes. Furthermore, by learning different positions of the same shape, the generative model can be used to interpolate between two positions and thus, animate shapes in a controlled direction.  

Another possible application of our method employs a geometric texture transfer using a \textit{two-step} mapping. First, we build a local geometric texture on a simple shape such as a sphere or a torus, where automatic or semi-automatic tools work well. Then, this mesh is used as an intermediate shape toward the ultimate goal of generating textures on the target mesh.
This two-step method is reminiscent of the 30+ year old work of \citet{bier1986two} for texture mapping. 

Learning to synthesize is a challenging task, especially when it comes to irregular geometric data. The proposed face convolution facilitates the development of a GAN framework for triangular meshes. Our learning-based technique leads to results that are difficult to achieve using state-of-the-art graphics tools, or requires a tailored solution for each target shape. We believe that this work is just a first step towards the development of more deep-learning techniques for 3D generative mesh models.
\begin{acks}
We would like to thank the anonymous reviewers for their helpful comments. This work is supported by the NSF-BSF grant (No. 2017729), the European research council (ERC-StG 757497 PI Giryes), ISF grant 2366/16, and the Israel Science Foundation ISF-NSFC joint program grant number 2217/15, 2472/17.
\end{acks}

\bibliographystyle{ACM-Reference-Format}
\bibliography{bibs}

\end{document}